\documentclass[acmsmall]{acmart}

\usepackage[utf8]{inputenc} %
\usepackage[T1]{fontenc}

\usepackage{wrapfig}

\usepackage[absolute,showboxes]{textpos}
\usepackage{multirow}
\usepackage{booktabs}
\usepackage{tabularx}
\usepackage{xcolor}
\usepackage{arydshln}
\usepackage{todonotes}
\usepackage[group-separator={,}]{siunitx}
\usepackage[labelformat=simple]{subcaption}

\usetikzlibrary{patterns,decorations.pathreplacing,arrows.meta,shapes.symbols,shapes.geometric,matrix,fit}
\usepackage{rotating}

\setcopyright{cc}
\setcctype{by}
\acmJournal{PACMNET}
\acmYear{2025} \acmVolume{3} \acmNumber{CoNEXT4} \acmArticle{50}
\acmMonth{12} \acmDOI{10.1145/3768997}



\received{June 2025}
\received[accepted]{September 2025}

\newcommand{\result}[1]{}

\definecolor{myred}{cmyk}{0, 0.7808, 0.4429, 0.1412}

\newcommand{\done}[1]{}

\definecolor{positive}{HTML}{3C8031}
\definecolor{negative}{HTML}{AF3235} %
\definecolor{neutral}{HTML}{E1A205} %

\usepackage{pifont}

\usepackage{xspace}
\newcommand{\etal}{\textit{et al.}~}
\newcommand{\eg}{\textit{e.g.,}~}
\newcommand{\ie}{\textit{i.e.,}~}

\newcommand{\one}{({\em i})\xspace}
\newcommand{\two}{({\em ii})\xspace}
\newcommand{\three}{({\em iii})\xspace}
\newcommand{\four}{({\em iv})\xspace}

\makeatletter
\renewcommand{\paragraph}[1]{\vspace{0.03in}\noindent{\bf #1.}\hspace{0.25ex \@plus1ex \@minus.2ex}}
\newcommand{\paragraphNoDot}[1]{\vspace*{0.03in}\noindent{\bf #1}\hspace{0.25ex \@plus1ex \@minus.2ex}}
\makeatother

\newcommand*\dhline{\specialrule{0pt}{1pt}{0pt}\hdashline[.4pt/3pt]\specialrule{0pt}{0pt}{2pt}}

\begin{document}

\title{Scanning the IPv6 Internet Using Subnet-Router Anycast Probing}

\author{Maynard Koch}
\orcid{0009-0009-3698-1342}
\affiliation{%
	\institution{TU Dresden}
	\city{Dresden}
	\country{Germany}
}
\email{maynard.koch@tu-dresden.de}

\author{Raphael Hiesgen}
\orcid{0000-0002-1676-8108}
\affiliation{%
	\institution{HAW Hamburg}
	\city{Hamburg}
	\country{Germany}}
\email{raphael.hiesgen@haw-hamburg.de}

\author{Marcin Nawrocki}
\orcid{0000-0002-6308-5502}
\affiliation{%
	\institution{NETSCOUT}
	\city{Westford}
  \state{MA}
	\country{USA}}
\email{marcin.nawrocki@netscout.com}

\author{Thomas C. Schmidt}
\orcid{0000-0002-0956-7885}
\affiliation{%
	\institution{HAW Hamburg}
	\city{Hamburg}
	\country{Germany}}
\email{t.schmidt@haw-hamburg.de}

\author{Matthias W\"ahlisch}
\orcid{0000-0002-3825-2807}
\affiliation{%
	\institution{TU Dresden}
	\city{Dresden}
	\country{Germany}
}
\email{m.waehlisch@tu-dresden.de}

\definecolor{boxgray}{rgb}{0.93,0.93,0.93}
 \textblockcolor{boxgray}
 \setlength{\TPboxrulesize}{0.7pt}
 \setlength{\TPHorizModule}{\paperwidth}
 \setlength{\TPVertModule}{\paperheight}
 \TPMargin{5pt}
 \begin{textblock}{0.8}(0.1,0.04)
   \noindent
   \footnotesize
   If you refer to this paper, please cite the peer-reviewed publication:
   Maynard Koch, Raphael Hiesgen, Marcin Nawrocki, Thomas C. Schmidt, and Matthias Wählisch. 2025.
   Scanning the IPv6 Internet Using Subnet-Router Anycast Probing.
 In \emph{Proceedings of the ACM on Networking (PACMNET) 3, CoNEXT4, Article 50 (December 2025)}. https://doi.org/10.1145/3768997
\end{textblock}

\begin{abstract}
Identifying active IPv6 addresses is challenging.
Various methods emerged to master the measurement challenge in this huge address space, including hitlists, new probing techniques, and AI-generated target lists.    
In this paper, we apply active Subnet-Router anycast (SRA) probing, a commonly unused method to explore the IPv6 address space.
We compare our results with lists of active IPv6 nodes obtained from prior methods and with random probing.
Our findings indicate that probing an SRA address reveals on average 10\%~more router IP addresses than random probing and is far less affected by ICMP rate limiting.
Compared to targeting router addresses directly, SRA probing discovers 80\% more addresses.
We conclude that SRA probing is an important addition to the IPv6 measurement toolbox and may improve the stability of results significantly.
We also find evidence that some active scans can cause harmful conditions in current IPv6 deployments, which we started to fix in collaboration with network operators.

\end{abstract}

\maketitle

\section{Introduction}
\label{sec:introduction}

Internet research often relies on measurements, which in turn require a representative picture of its population for interpretation. In IPv4, researchers commonly make use of probing the entire address space, which is quickly achievable thanks to stateless scanning~\cite{dwh-zfiws-13}. The huge IPv6 address space renders related approaches impossible, leaving measurement researchers with a largely unknown object of investigation. 

About a decade ago, the community started to counter this restriction by collecting active IPv6 addresses~\cite{ukkw-ripsa-15}. 
Since 2016, the TU Munich (TUM) assembles a comprehensive hitlist of active IPv6 nodes~\cite{gsgc-siitc-16}, which established as a highly valuable community service. Since then, a significant research body has built upon the TUM Hitlist as input to various IPv6 measurement studies. 
However, scalable IPv6 scanning that does not rely on external input sources other than BGP announcements remains an unsolved challenge.

In this paper, we reconsider state-of-the-art IPv6 scanning approaches and propose a rarely used method, Subnet-Router anycast (SRA) probing, to explore the IPv6 address space.
In SRA probing, a scanner sends packets to an IPv6 address that represents a potential subnet prefix and all host bits are set to zero, \eg the SRA address of \texttt{2001:db8:1::/48} is \texttt{2001:db8:1::}.
Unlike probing random addresses, SRA probing significantly reduces the impact of ICMPv6 error message rate limiting, as this method triggers ICMPv6 Echo replies from routers operating interfaces directly connecting the probed subnets.

The remainder of this paper is structured as follows.
\vspace{-0.10cm}
\begin{enumerate}
	\item We introduce Subnet-Router anycast probing and partition the routable IPv6 address space into smaller subnets (\S~\ref{sec:method}).
  \item We provide detailed insights into our measurement results (\S~\ref{sec:sra-measure}). Our findings show that SRA probing finds on average 10\% more addresses than random probing for routers in active subnets. SRA probing discovers 80\% more addresses compared to targeting router addresses directly. We observe varying response rates for ICMPv6 Echo Reply messages between 1.3\% and 35.2\% depending on the targeted subnet set.
	\item We compare our SRA probing results to popular public datasets as well as passively collected IXP flowdata (\S~\ref{sec:sra-comparison}). The majority (97\%-99.9\%) of IPv6 addresses we discover with SRA probing are not contained in any of the datasets we use for comparison.
	\item We discover a critical amplification threat related to scanning the current IPv6~infrastructure (\S~\ref{sec:loop-amples}) and report about mitigation with operators. 
\end{enumerate}
 
We conclude from our findings that our current perspective on the operational IPv6 Internet is still largely incomplete and further efforts are needed to overcome our limited view (\S~\ref{sec:discussion}).

\section{Background and Related Work}
\label{sec:background}

\begin{figure}[b]
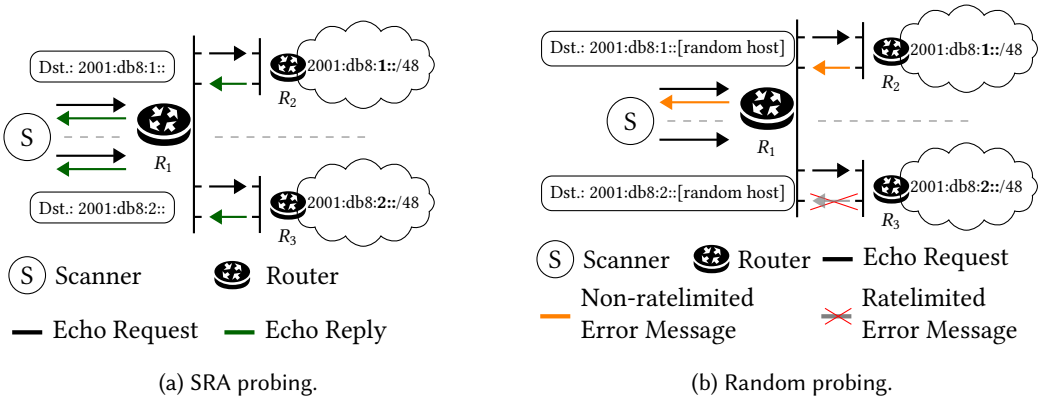

  \begin{subfigure}{0.45\textwidth}
    \input{figures/sra-probing-overview.tex}
    \caption{SRA probing.}
  \label{fig:sra-probing}
  \end{subfigure}
  \hfill
  \begin{subfigure}{0.49\textwidth}
    \input{figures/random-probing-overview.tex}
    \caption{Random probing.}
  \label{fig:rand-probing}
  \end{subfigure}
  \caption{IPv6 Scanning based on different probing methods. Random probing leads to ICMP error message rate~limiting (at $R_3$), therefore we discover more router IP addresses with SRA probing because SRA elicits ICMP Echo replies instead of ICMP error messages.}
  \label{fig:sra-vs-random-probing}
\end{figure}
\setlength{\tabcolsep}{2.75pt}
\begin{table*}
	\centering
	\caption{Overview of active and passive IPv6 measurement methods considered in this work.}
	\label{tbl:methods-and-bias-overview}
	\begin{tabular}{lllr}
		\toprule
    Method & Discovery of & Observed Addresses [\#]\\
		\midrule
    Random Probing & Router (Core) & 1.3M~\cite{bdpr-ibsai-18}, 9.5M~\cite{ipv6_allpref_topology_102024} \\
		& Router (Periphery) & 64M~\cite{rb-dtnp-20}, 52M~\cite{llzdl-finpd-21}, 44M~\cite{hmu-drwie-24} \\
		Hitlist &  Active End Hosts & 20M~\cite{ipv6-hitlist-service}\\
    IXP Flows & Active End Hosts & 146M~\cite{gsgc-siitc-16}, 198M (this work)\\
    \dhline 
		SRA Probing (this work) &  Router (Core and Periphery) & 133M\\
		\bottomrule	
	\end{tabular}
\end{table*}

\paragraph{Subnet-Router anycast addresses}
Subnet-Router anycast (SRA) addresses have been originally introduced in RFC~1884~\cite{RFC-1884} in 1995.
Since then, they are integral part of IPv6~\cite[\S 2.6.1]{RFC-4291}.
SRA~addresses enable applications to communicate with a router of the subnet, without knowing the actual IPv6 router address.

Syntactically, SRA addresses are unicast addresses.
They represent the subnet, \ie the host part of the IPv6 address is set to 0, \eg for \texttt{2001:db8:1::/48} the SRA address is \texttt{2001:db8:1::0}.
Each router is required to support a Subnet-Router anycast address if it has an interface to this subnet.
Routers receiving a packet targeting an SRA address of one of its subnets should reply with their own full source address.
For example, a router with the two interfaces \texttt{2001:db8:1::2/48} and \texttt{2001:db8:10::2/48} receives a packet to \texttt{2001:db8:1::0} via its interface \texttt{2001:db8:10::2/48} will reply with the source address \texttt{2001:db8:10::2}.
Notably, a router is not required to reply, there are behavioral differences of SRA implementations~\cite{s-bdisr-23}.
We illustrate SRA probing in \autoref{fig:sra-probing}, in contrast to random probing (see \autoref{fig:rand-probing}) we circumvent ICMP error message rate limiting at $R_1$, thus we discover more router addresses with SRA probing.
We will make use of SRA~addresses to explore new active subnets and new IPv6~router addresses.
So far, only very limited insights into this approach exist \cite{b-nrirb-20}, and we provide the first comprehensive measurements.

\paragraph{Active measurements}
A full scan of the IPv6~address space is not scalable.
Therefore, active measurements, which send probe packets to specific target addresses, try to limit the scope of targets.
A straightforward option is to send one probe packet to any IPv6~prefix visible in public BGP dumps~\cite{ipv6_allpref_topology_102024,ripe-atlas,yc-eirid-24,ehksw-dmvis-25}.
This reduces probing (\ie ~addresses) but is relatively selective since BGP announcements include larger covering prefixes of more specific active subnets.
To broaden coverage, prefixes available in BGP are split into equally sized, more specific prefixes, usually of size \texttt{/48} or~\texttt{/64}.

When deciding on the actual target address, there are currently three principal options.
\one Select a random address within a given (sub)prefix.
\two Select an address from a hitlist.
\three Artificially generate addresses by extrapolating knowledge about assignment of IPv6~addresses.

Targeting a random address has very high chances of not reaching an active end host address, instead replies rather reveal routers. %
Holzbauer \etal~\cite{hmu-drwie-24} show that related ICMP error messages issued by routers can be used to get a better understanding of active networks.
Selecting an address from a hitlist is often combined with traceroute measurements to discover nodes (\ie router, middleboxes) between source and destination.
Tools such as ZMapv6~\cite{g-zisic-22}, XMap~\cite{llzdl-finpd-21}, and Yarrp~\cite{b-yirhs-16,g-hfmii-17} provide those capabilities for IPv6.
Low-byte addresses such as \texttt{<Prefix>::1} reflect a common assignment pattern.
CAIDA Ark~\cite{ipv6_allpref_topology_102024} and RIPE Atlas~\cite{ripe-atlas} are distributed measurement platforms that regularly target low-byte addresses.
Target Generation Algorithms (TGA) try to discover more advanced structures to predict other potentially active addresses~\cite{ukkw-ripsa-15,fpb-eiusi-16}.
Steger \etal~\cite{skzcg-taetg-23} evaluated the hit rate of TGAs compared to the TUM Hitlist.
One fundamental challenge of those approaches is that they require seeding, usually based on knowledge gathered via passive or other active measurements---the output depends on the input.
We give an overview of active probing techniques and the number of discovered addresses in \autoref{tbl:methods-and-bias-overview}.

\paragraph{Challenges in active measurements}
Active scans face the challenge of ICMP rate limiting~\cite{rub-cirlr-15,pyhwn-yrpmi-22} since ICMPv6~\cite{RFC-4443} requires that a node must limit the rate of ICMPv6 error messages it originates.
Such rate limiting can cause irregular on-off behavior of routers~\cite{rub-cirlr-15}, therefore, impacting the reliability and stability of scan results.
This harms not only scans targeting locally unreachable networks (``No Route to Destination'') but, in particular, random addresses (``Address Unreachable'').
We tackle this challenge by targeting Subnet-Router addresses---addresses to which one router should always reply with an ICMPv6 Echo instead of an error when the network is~active.

\paragraph{Passive measurements}
Passive measurements derive active IPv6~addresses by observing IPv6 communication or finding additional hints that refer to potentially active IPv6~addresses.
Those approaches include the analysis of server logs of common services such as NTP pools~\cite{rl-ihsbc-23}, or exploit non-existing \cite{fbhkv-sfntc-17} or existing~\cite{bdpr-ibsai-18,rb-dtnp-20,wehbb-6iiss-24} IPv6 DNS records based on reverse DNS or common names, \eg derived from top lists~\cite{shgjj-lwtss-18} or TLS~certificates publicly available in CT logs~\cite{gsgc-siitc-16}.

Those and other sources are compiled in the TUM Hitlist~\cite{gsgc-siitc-16,zssgc-rcdir-22,skzcg-taetg-23}, the most popular public list of active IPv6~addresses, which serves in many studies as a point of reference (\eg \cite{syhwl-acgai-22,ychz-6elat-22,hcwyz-6hdii-23,wehbb-6iiss-24}).
The initial data set of the 2016~TUM Hitlist contained IPv6~addresses based on passive flow data captured at a large Internet Exchange Point. Their IXP collection comprises 146M unique IPv6 addresses (see \autoref{tbl:methods-and-bias-overview})

In our study, we compare the addresses contained in the TUM Hitlist and our one-month address collection from the IXP with the addresses we discover through SRA probing (see \autoref{tbl:methods-and-bias-overview}).

\section{Measurement Method and Setup}
\label{sec:method}

We propose Subnet-Router anycast probing, an IPv6-compliant approach using ICMPv6 Echo requests. 
This active, controlled measurement aims for detecting routers at the Internet core and edge, depending on the probed subnets. 
Little to no attention has been paid to scanning SRA addresses to unveil IPv6 router infrastructure.

Prior active methods to explore the IPv6 space~\cite{b-yirhs-16,bdpr-ibsai-18,yc-eirid-24,rb-dtnp-20,llzdl-finpd-21,hmu-drwie-24} rely on random probing of the IPv6 address space.
While targeting unassigned addresses leads to ICMPv6 error messages, these messages only give an incomplete picture due to ICMPv6 error message rate limits~\cite{RFC-4443}.
We instrument SRA addresses to mitigate rate limiting in our measurements since requesting these addresses triggers ICMPv6 Echo replies.
We now describe our method and setup to deploy large-scale SRA scans.

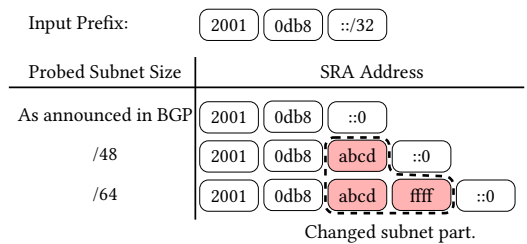
\begin{wrapfigure}{r}{0.5\textwidth}
  \vspace{-30pt}
  \resizebox{0.5\textwidth}{!}{\begin{tikzpicture}[
  box/.style={draw,rectangle, rounded corners, minimum height=5mm, minimum width=9.5mm, font=\myfontsize},
  pinkbox/.style={box, fill=red!30}
]
\newcommand{\myfontsize}{\small}
\node[align=left] (label1) {\myfontsize Input Prefix:};
\matrix[matrix of nodes,
        nodes={box, anchor=center},
        column sep=0.75mm,
        row sep=4mm,
        right=0.9cm of label1] (prefix) {
  2001 & 0db8 & ::/32 \\
};

\node[below=.75cm of label1.west,anchor=west] (label2) {\myfontsize Probed Subnet Size};
\node[right=2cm of label2.east] (label3) {\myfontsize SRA Address};

\node[below=.45cm of label2.center,align=center] (label4) {\myfontsize As announced in BGP};
\node[below=.4cm of label4.center] (label5) {\myfontsize /48};
\node[below=.4cm of label5.center] (label6) {\myfontsize /64};
\matrix[matrix of nodes,
        nodes={box},
        column sep=0.75mm,
        row sep=1mm,
        below=2.1cm of prefix.west,anchor=west] (tbl) {
  2001 & 0db8 & ::0 \\
  2001 & 0db8 & |[pinkbox] (p1)| abcd & ::0 \\
  2001 & 0db8 & |[pinkbox] (p2)| abcd & |[pinkbox] (p3)| ffff & ::0 \\
};

\draw[thick] ([yshift=-0.25cm,xshift=-0.2cm]label2.west) -- ([yshift=-0.25cm,xshift=1.5cm]label3.east);
\draw[thick] ([xshift=0.1cm]label2.north east) -- ++(0,-2.6);

\draw[dashed, rounded corners,very thick]
  ([xshift=-1pt,yshift=1pt]p1.north west) --
  ([xshift=1pt,yshift=1pt]p1.north east) --
  ([xshift=1pt,yshift=-1pt]p1.south east) --
  ([xshift=1pt,yshift=1pt]p3.north east) --
  ([xshift=1pt,yshift=-1pt]p3.south east) --
  ([xshift=-1pt,yshift=-1pt]p3.south west) --
  ([xshift=-1pt,yshift=-1pt]p2.south west) --
  cycle;

\node[below=0.35cm of p2.east,xshift=0.05cm] (label7) {\myfontsize Changed subnet part.};

\end{tikzpicture}}
    \caption{Example construction of a single SRA address for every target subnet given a single input prefix.}
  \label{fig:sra-construction}
  \vspace{-10pt}
\end{wrapfigure}
\subsection{Subnet-Router Anycast Probing}
Querying Subnet-Router anycast (SRA) addresses should trigger an ICMPv6 Echo reply from a router that has an interface for the respective subnet.
To successfully deploy our probing, we need to partition the IPv6 address space such that each partition represents an active subnet, cope with IPv6 aliasing, and map a reply to the original request since the source address of the replying router may be unrelated to the targeted subnet.

\paragraph{Partitioning the address space announced in BGP to create SRA addresses}
At the time of writing, approximately 200k~IPv6 prefixes are announced in BGP.
Probing the Subnet-Router anycast address of each routable~prefix as it is announced in BGP misses internal, more specific subnets.
To balance scan traffic and increase the chance of discovering new addresses, we partition the routable address space into three stages. We start by querying the SRA address of each announced prefix (Stage 1). Then, we partition the routable address space into /48 subnets (Stage 2) and scan the SRA address of each subnet.
Even though common routing policies should prevent announcements more specific than a /48, we found 3k prefixes more specific than /48.
In these cases, we scan the SRA address of the /48~supernet, unless it is included in another announcement. Partitioning into /48 subnets results in ~15 billion potential targets. Finally, we partition all /48 announcements ($\approx$100k) further into /64 subnets (Stage~3). This step generates +$\approx$6.8 billion target addresses for scanning. 

For each stage, we construct the target addresses as follows. In Stage 1, we leave all bits of the given input prefix unchanged. In Stage~2, we create all bit combinations of the first 16-bit~block that follows the original subnet prefix, which yields $2^{16}$ new addresses per input prefix.
In Stage 3, we take the subsequent 16-bit block into account and generate all possible $2^{32}$ combinations to construct SRA addresses for probing.
As the number of addresses to probe grows exponentially, we limit the third stage to only use \texttt{/48} announcements and do not generate addresses more specific than a \texttt{/64}.
We employ this multi-stage approach to determine which input set is most suitable for SRA probing.
\autoref{fig:sra-construction} shows an example address for each stage given the input prefix~\texttt{2001:db8::/32}.

\setlength{\tabcolsep}{2.75pt}
\begin{table*}[t]
	\centering
\caption{Comparison of different input sets to probe Subnet-Router anycast addresses and their effectiveness for SRA probing. The hitlist input reveals the most router IP addresses while probing only 700M SRA addresses.}
	\label{tbl:results:datacorpus}
	\begin{tabular}{lrrrrr}
		\toprule
		\multicolumn{4}{c}{Input for probing} & \multicolumn{2}{c}{Results} 
		\\
		\cmidrule{1-4}
		\cmidrule(l){5-6}
		Source & Subnets [\#] & Subnet-size & Addr.  [\#] & Replies [\#] & Router IPs [\#]\\
		\midrule
		BGP (All) & 200k & As announced & 200k & 38k (19\%)& 28k (14\%)\\
		BGP (All) & 200k & /48 & 11B & 350M (3.2\%)& 4M (0.04\%)\\
		Route(6) (All) & 1M & /64 & 10B & 570M (5.7\%)& 14M (0.14\%)\\
		BGP (/48) & 100k & /64 & 6.5B & 1.3B (20\%)& 45M (0.69\%)\\
		Hitlist (Unique /64s)& 700M& /64 & 700M & 90M (13\%) & 72M (10.3\%)\\
		\midrule
		Total & & & 28.2B & 2.32B & 135M\\
		& & & &  & (Distinct 133M)\\
		\bottomrule
	\end{tabular}
  \vspace{-10pt}
\end{table*}

\paragraph{Creating SRA addresses using other input sources}
BGP announcements reflect intended reachability. There are, however, other sources containing more specific subnet assignments, which can be used to leverage the effectiveness of our method.
In addition to BGP announcements, we consider two input sources to create SRA addresses.
First, we collect \texttt{Route(6)} objects from IRR databases, which predominantly contain /48 prefixes. For each of the nearly 1M prefixes, we create up to 10k random /64 SRA~addresses, adding up to 10B targets.
Second, we construct a target set from the TUM Hitlist (2.5B~addresses) by taking the first 64 bits of each host address and set the remaining 64 bits to zero, which results in 700M distinct targets.

\paragraph{IPv6 Alias Resolution}
In IPv6, operators may configure their networks as aliased, which means they reply to, \eg ICMPv6 Echo requests on any address. Several studies have investigated this effect and developed methods to detect if a subnet is aliased or not~\cite{lbbc-sisia-13,gsflk-ceuui-18}. To overcome this issue in our measurements, we make use of the fact that SRA addresses are typically not assigned to hosts; therefore, we filter for replies that originate from the same source we requested (the \texttt{::0} address), as this is an indicator for an aliased network. Additionally, we check whether the remaining source IP addresses are part of the aliased prefix list provided by the IPv6 hitlist service~\cite{ipv6-hitlist-service}. Our approach reduces the impact of aliased prefixes and serves as a trade-off to maintain high scan performance. We are aware of the limitation that this approach may misclassify a small portion of addresses, but we consider the impact to be negligible. 

\paragraph{Capturing replies}
We need to match the target IP~address in our probing packet (\ie the SRA address) to the replying IP address (\ie the router IP~address).
To that end, we encode the target SRA~address in the ICMPv6 payload and extract it from the incoming reply. 
This takes advantage of ICMP, which includes parts of the original request in the reply. 
In addition to ICMPv6 Echo replies, we capture any type of ICMPv6 error message such as ``Time Exceeded'' or ``No Route to~Host''.

\paragraph{Activity of router addresses}
Hitlists focus on active host discovery. %
To that end, we inspect the activity of all found addresses--from both, ICMPv6 Echo replies and ICMPv6 error messages--by probing them once a day over the course of a week from November 5 to November 13, 2024. We report on our findings in \autoref{sec:sra-measure}.

\paragraph{Stability of SRA probing}
To better understand the effectiveness of SRA probing, we inspect if re-probing the SRA address elicits a response from the same router IP address. While changes may indicate a network change, they would significantly impact the reliability of results obtained from SRA probing. Therefore, we re-probe the \texttt{/64} SRA addresses we created based on the TUM Hitlist in six scans, distributed over two days, and analyze the stability of the found addresses.

\subsection{Measurement Setup}
We perform our active probing in October and November~2024 and repeat one SRA BGP (only \texttt{/48}) scan in February~2025.
When using input sources other than BGP, \eg the TUM Hitlist, we ensure that datasets are aligned in time.
We use a single vantage point located in Europe, connected to a large IXP and with 1Gbit/s upstream. %
Neither we nor our upstream filters traffic.
For active probing, we use a fork of the TU Munich ZMapv6 tool~\cite{g-zisic-22} with a modified version of the ICMPv6 Echo scan module and a custom GO implementation as an address generation tool, serving as input for ZMap. We limit the scan rate to 200k packets per second.
We perform each scan at least twice, the re-probing of the found router IP addresses is done seven times, and we run SRA probing of the TUM Hitlist \texttt{/64} six times. The final list of router IP addresses is compiled from the initial scan of each input source.
\begin{wrapfigure}{r}{0.5\textwidth}
  \vspace{-10pt}
    \includegraphics[width=0.5\textwidth]{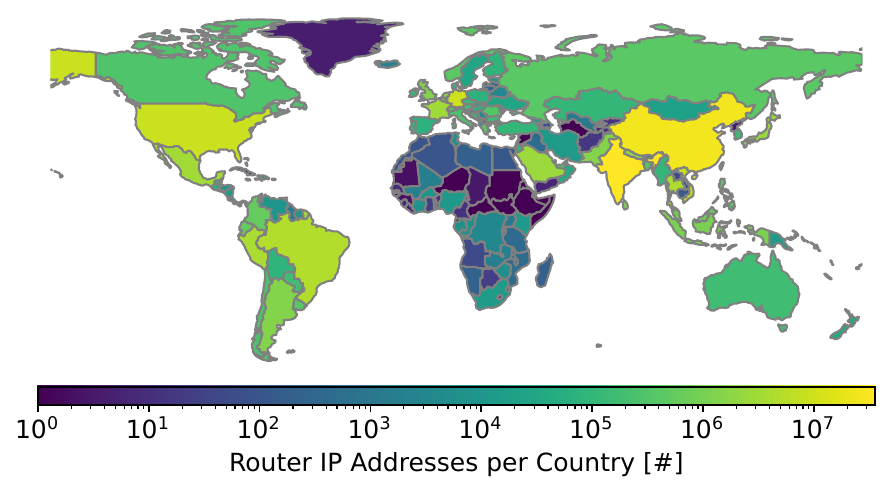}
  \caption{World-wide distribution of router IP addresses found with SRA probing.}
    \label{fig:worldmap}
  \vspace{-20pt}
\end{wrapfigure}
\paragraph{Metadata mapping}
We use the free MaxMind GeoIP database~\cite{maxmind-geolitecountry} to map source IP addresses to countries, the RouteViews dataset~\cite{routeviews} to map IPv6 addresses to Autonomous System Numbers~(ASNs), and the IPinfo ASN database~\cite{ipinfo-asndb} to identify the types of Autonomous Systems (ASes) associated to router IP addresses we reveal with SRA~probing.

\section{Subnet-Router Anycast Measurements}
\label{sec:sra-measure}
\begin{figure}[t]
	\includegraphics[width=.8\linewidth]{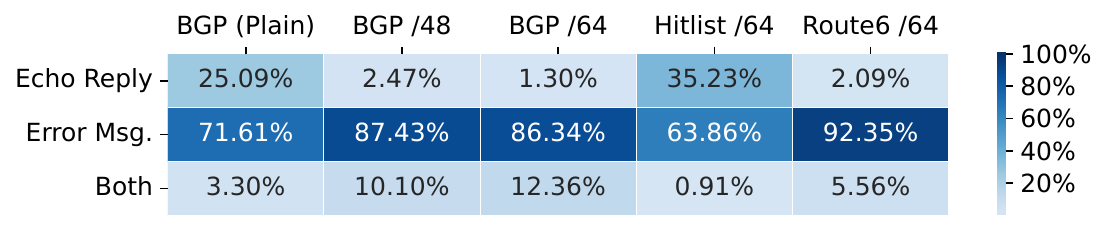}
	\caption{Relative ratio of ICMP replies, grouped into Echo replies, error messages, and ambiguous values for router IP addresses that sent error messages for some probed subnets and ICMPv6 Echo replies for~others.}
	\label{fig:results:icmp-replies-allscans}
  \vspace{-5pt}
\end{figure}
We probe more than 28B Subnet-Router anycast addresses in 5 scans, observing more than 2.3B~replies from 133M distinct IP addresses distributed in over 218 countries (see \autoref{fig:worldmap}) with a strong bias towards India (27\%) and China (20\%). We observe widely varying response rates per scan, ranging from 3.2\% (BGP /48) up to 20\% (BGP /64). The ratio between newly discovered routers and probed IP addresses strongly depends on the input source, see \autoref{tbl:results:datacorpus}. With 72M unique router IP addresses discovered by sending only 700M requests (all unique /64s generated from the full TUM hitlist) the discovery rate exceeds 10\% while it remains below 1\% for all other scans, except for the plain BGP prefix scan, which shows a higher reply rate in relative terms. With only 28K router IP addresses and an overlap of >90\% with the other scans, the effect is negligible.
The reason for that is that the different data sources provide different types of networks. Generating more specific subnets based on BGP (or \texttt{Route(6)}) input data, usually leads to subnets that are not assigned to any router interface, triggering many ICMP error messages when probed.
On the other hand, the /64 subnets resulting from the TUM Hitlist are not artificially generated but cut-off from the host address of an (at least at some point in the past) active host.
Therefore, it is much more likely that the probed subnet is active and assigned to a periphery router, which throws no error message but responds with an ICMPv6 Echo~reply.

\paragraph{ICMP Response Types}
\autoref{fig:results:icmp-replies-allscans} illustrates the per scan distribution of Echo replies vs. error messages. While we receive more error messages in general, the \texttt{Route(6)} /64 scan receives more than 92\% error messages for probed subnets. The reason for that lies in the random generation of 10k /64 subnets. Nearly 50\% of the \texttt{Route(6)} objects announce a /48.
This means that the random subnet generation covers only 15\% of the possible /64 subnets for a single /48. Our observations also imply that these subnets are sparsely populated. We observe the highest Echo reply rates~(35\%) for the Hitlist /64 scan---probing the periphery of a network is far more effective in case of SRA~probing.
This confirms our prior assumption that deriving SRA~addresses from active devices increases chances to hit a subnet router, which ultimately leads to significantly more Echo~replies.

\paragraph{Advantage of SRA probing}
SRA probing shows full advantage when re-probing active subnets.
Consecutive scans are not affected by common ICMP error message rate limiting~\cite{rub-cirlr-15,pyhwn-yrpmi-22} since sending probes to an enabled SRA~address will trigger an ICMP Echo reply message, independently whether the router IP~address changes or not.
Random probing, on the other hand, triggers ICMP~error messages (\eg ``Address Unreachable'') when a successfully probed random address changes in the future, and too many error messages will be~suppressed.
Additionally, the chance to hit an active device at all using a random IPv6 address is almost zero.

\begin{figure}%
	\includegraphics[width=.8\textwidth]{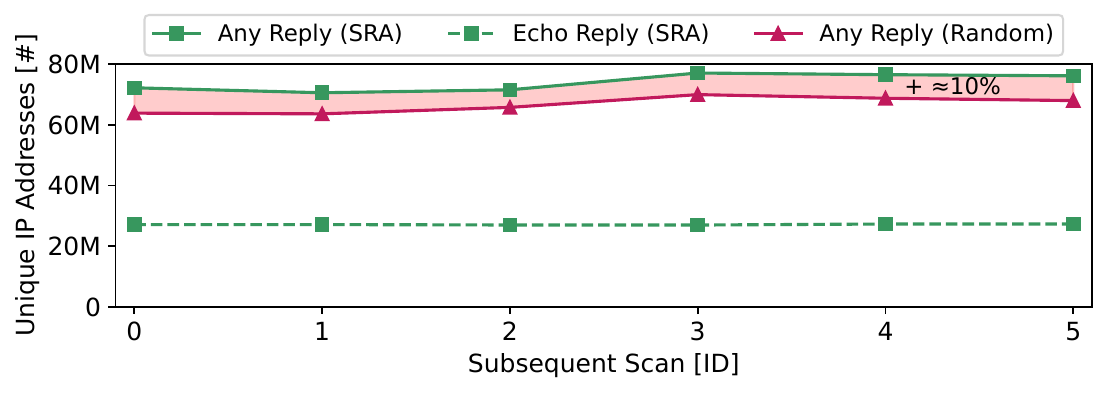}
	\caption{Comparison of SRA vs. random probing of all /64s from the TUM Hitlist. With SRA probing, we observe $\approx 10\%$ more addresses than with random probing. While the total number of replies varies, the number of Echo replies remains stable.}
	\label{fig:results:tum64-sra-vs-random}
  \vspace{-20pt}
\end{figure}

\autoref{fig:results:tum64-sra-vs-random} shows the total number of discovered router IP addresses based on SRA and random probing for a measurement series using the Hitlist /64 subnets.
Per scan campaign, we find about 10\% more router IP addresses with SRA probing. The number of router IP addresses that respond with an Echo reply message remains stable, which clearly shows that our SRA scans are far less affected by ICMP rate limiting. We also find $\approx 9M$ router IP addresses exclusively with SRA probing, which strengthens the use of SRA probing.
The overlap of two consecutive scans, however, remains at a steady level but is below 70\%. The observability of a router IP address is highly influenced by ICMPv6 rate limiting.

\begin{figure}%
  \begin{subfigure}{0.49\textwidth}
    \includegraphics[width=1\textwidth]{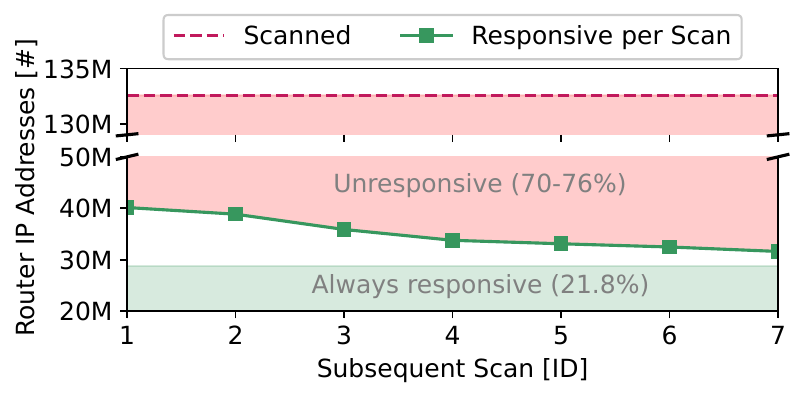}
    \caption{Visibility, \ie probing the router IP addresses directly. 28M router IP addresses replied to Echo requests in any scan.}
    \label{fig:results:routerstability}
  \end{subfigure}
  \hfill
  \begin{subfigure}{0.49\textwidth}
    \includegraphics[width=1\textwidth]{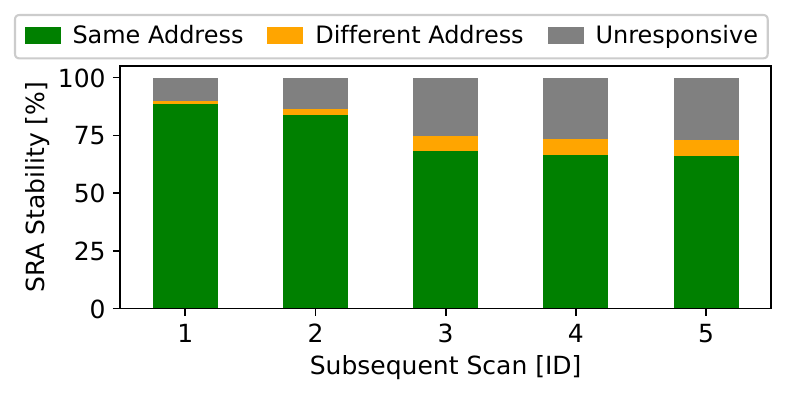}
    \caption{Stability, \ie re-probing SRA addresses. For at least 66\%, the same SRA address triggered a reply from the same router IP address in all scans.}
    \label{fig:results:srastability}
  \end{subfigure}
  \caption{Visibility and stability of the discovered router IP addresses.}
  \vspace{-15pt}
\end{figure}

\paragraph{Visibility and stability of router addresses}
Stability of addresses is an important factor when creating hitlists.
To better understand the potential of SRA scanning in finding router addresses that remain reachable, we analyze if we \one elicit a response when probing the router IP address directly and \two trigger a response from the same router IP address when re-probing the same SRA~address. We re-probe all router addresses found in each SRA~scan every day for a week and re-probe the Hitlist /64 SRA addresses six times within 2 days.
\autoref{fig:results:routerstability} illustrates the churn of detected router IP addresses. Only 28M out of 133M addresses always reply to Echo requests. However, we observe a significantly higher stability when re-probing the SRA address (see \autoref{fig:results:srastability}). After two days, we observe 66\% of the probed SRA addresses to still reveal the same router IP address as before, changes are rare (max. 7\%), and we do not elicit a response for the same SRA address for about 27\% in the last scan. Re-probing after three months revealed that while for 18\% the SRA address reveals another router IP address, about 40\% of the router IP addresses are still reachable via the same SRA~address. These results show that most routers do not reply to direct ICMPv6 Echo requests but are to a major part reliably discoverable through SRA probing.

\paragraph{Prevalence and stability of ASNs and IPv6 prefixes}
The top 5 ASes from which the router IP~addresses originate are shown in \autoref{tbl:asn-allsources}. The router addresses we find (first group of columns) cover 16k~ASNs, and 11\% of the IP addresses originate from the top ASN. We observe a clear bias towards networks in Asia, being more responsive than others.
Over the course of our six~consecutive scans (not shown), $\approx$~87\% of the prefixes remain unchanged, leading to a stable set of ASes of $\approx$~96\%. 
We further report on the network type distribution in \autoref{sec:network-types}.

\vspace{-5pt}
\section{Comparing SRA Probing to Other Datasets} 
\label{sec:sra-comparison}
We compare our measurement results with multiple datasets of different characteristics.
These datasets are
\one publicly available IPv6~traceroute measurements,
\two a popular public hitlist of active hosts, and
\three flow data of one month provided by a large regional Internet Exchange Point~(IXP).

We observe little overlap in terms of IP addresses, and each data source reveals a different set of ASNs from which most of the IP~addresses originate, signifying the diversity of all collected datasets (see \autoref{tbl:asn-allsources}).
Considering all ASNs, however, shows that more than 99\% of the ASes found through SRA~probing are also present in the other datasets (see \autoref{fig:results:overlapASN}).
We analyze these observations further in this section.
\subsection{Comparison with Public IPv6 Traceroute Measurements}
\label{sec:cmp:public-traceroute}

We use traceroute data provided by CAIDA~\cite{ipv6_allpref_topology_102024} and RIPE Atlas~\cite{ripe-atlas}.
Both datasets are from October~4,~2024 to align in time with our measurement campaign.
It is worth noting that we target orders of magnitude more IP~addresses (28.8B targets vs 17.7M~targets in the CAIDA dataset and 687k~targets in the RIPE dataset) and that the methods differ.
We argue that the comparison is still appropriate for the following reasons.
First, these datasets are common sources of comparison when analyzing IPv6~scanning methods.
Second, our measurements take 1.5~days to cover 28.2B~SRA~addresses.
The RIPE~Atlas measurements, for example, need 13~days.
Even if we probe more targets in our setup, we do not treat other methods unfairly given the number of addresses explored per time.
Neither Ark nor RIPE Atlas would be able to scan 28.8B targets, and we are interested in what is possible given the current state of the art and SRA~probing.

\begin{table}%
  \centering
  \caption{Top 5 ASes per data source and the relative share of IP addresses per AS. We highlight ASes from SRA probing that are among the top 5 ASes in at least one other data source (bold font).}
  \label{tbl:asn-allsources}
  \small
\begin{tabular}{lrrrrrrrrrr}
\toprule
& \multicolumn{4}{c}{This Work}  & \multicolumn{4}{c}{Traceroute Measurements} &\multicolumn{2}{c}{Other} \\
		\cmidrule(lr){2-5}
		\cmidrule(lr){6-9}
		\cmidrule(lr){10-11}
& \multicolumn{2}{c}{SRA Probing} & \multicolumn{2}{c}{IXP Flows} & \multicolumn{2}{c}{CAIDA ITDK} & \multicolumn{2}{c}{RIPE Atlas} & \multicolumn{2}{c}{TUM Hitlist} \\
		\cmidrule(lr){2-3}
		\cmidrule(lr){4-5}
		\cmidrule(lr){6-7}
		\cmidrule(lr){8-9}
		\cmidrule(lr){10-11}
  Rank & ASN & IP addr. [\%] & ASN & IP addr. [\%] & ASN & IP addr. [\%] & ASN & IP addr. [\%] & ASN & IP addr. [\%] \\
\midrule
  1 & \textbf{45609} & 11.15 & 6805 & 43.06&      36183 & 15.49 & 16509 & 1.93 & 13335 & 25.62\\
2 & \textbf{9808} & 6.07 &  3209 & 21.55 &     16509 & 11.18 & 11172 & 1.89 & 12322 & 9.97 \\
3 & 45271 & 4.63 & 8881 & 8.50  &     13335 & 6.04 & \textbf{9808} & 1.78 & 5607 & 3.38 \\
4 & 38266 & 4.27 & 16202 & 6.06 &     19551 & 2.96 & 174 & 1.58 & 55836 & 3.20 \\
5 & \textbf{4134} & 4.12 &  20880 & 4.23 &      \textbf{45609} & 2.83 & 1299 & 1.22 & \textbf{4134} & 1.99 \\
\bottomrule
\end{tabular}
\vspace{-15pt}
\end{table}
\begin{wrapfigure}{r}{0.5\textwidth}
  \includegraphics[width=0.5\textwidth]{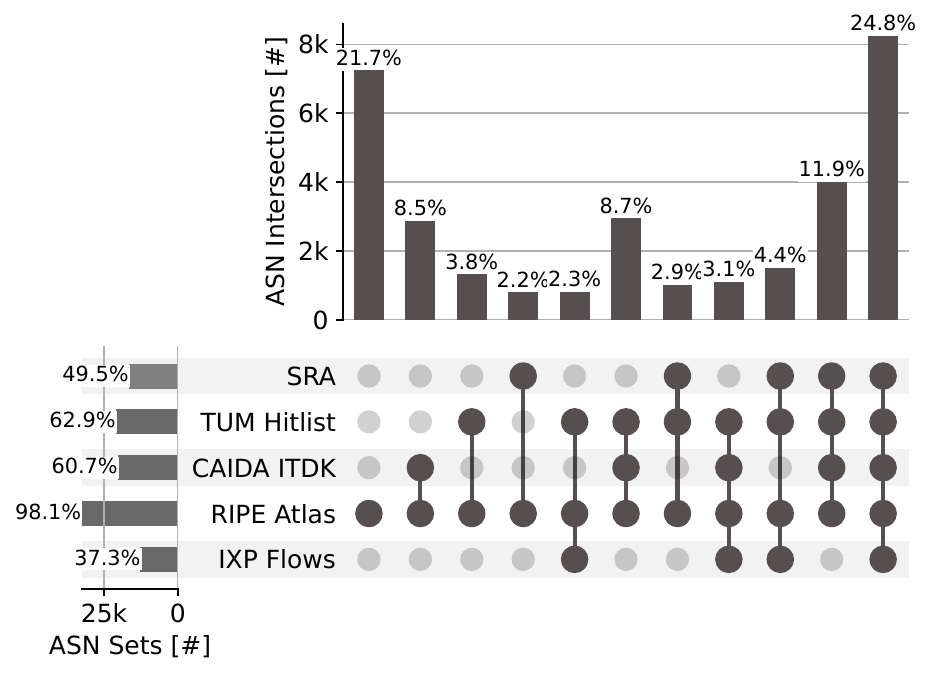}
  \caption{Overlap of ASes between all data sources. Subsets <2\% are not shown but presented in \autoref{sec:overlap-asn}.}
  \label{fig:results:overlapASN}
\vspace{-5pt}
\end{wrapfigure}
\paragraph{CAIDA Ark IPv6 Topology Dataset}
A globally distributed set of Archipelago (Ark) nodes continuously probes all IPv6 prefixes announced in BGP every 24~hours.
Each node sends traceroutes to a single random destination address as well as the address \texttt{<prefix>::1} of each prefix.
We discover 133M router IP addresses that are not included in the CAIDA data set, and CAIDA discovers 9.4M addresses we do not observe.
These observations further motivate the additional benefit of probing SRA addresses since they highly improve topology measurements with additional~data.

\paragraph{RIPE Atlas Traceroute Measurements}
RIPE Atlas is a measurement platform managed by the RIPE Network Coordination Centre (RIPE NCC) and consists of thousands of globally distributed nodes.
Similar to the CAIDA IPv6 Topology Dataset, RIPE Atlas~\cite{ripe-atlas-topology-msm} target the \texttt{<prefix>::1} address of each prefix announced in BGP.

We base our comparison on data collected by 6549 globally distributed Atlas nodes targeting 624k~IPv6 addresses.
The Atlas scans reveal 401k router IP~addresses, which is only 0.3\% of what the SRA probing discovered.
On the other hand, we discover only 48.5k router IP addresses that RIPE Atlas also does.
Similar to CAIDA, the overlap is minimal, highlighting the benefits of SRA address probing and the potential for combining methods to achieve more comprehensive datasets.
In terms of ASes, RIPE Atlas contains a large number of IP~addresses belonging to ASes that are not observed in any of the other data sources (see \autoref{fig:results:overlapASN}).
There are two reasons.
First, RIPE Atlas probes are deployed in much more ASes than probes of other traceroute projects.
Second, routers use different source IP~addresses when replying to SRA requests than to traceroute.
In the case of SRA, some routers use the peering LAN IP~address as source address, which often belongs to the IP~address space of the upstream provider, while replies to traceroutes tend to use the IP~address of the served subnet.
Mapping router IP~addresses to ASNs is therefore more error-prone in the case of SRA~probing.
\subsection{Comparison with a Public IPv6 Hitlist}
We rely on the TUM IPv6 hitlist service~\cite{ipv6-hitlist-service}.
This service regularly provides a list of active, dealiased IPv6 end-hosts.
We compare it to the extended version of the hitlist, which also includes the results of traceroute measurements targeting the collected active hosts, in order to find additional router addresses~\cite{gsgc-siitc-16}.
We use the hitlist from September 21, 2024, which contains approximately 20 million active hosts.
We discover 4.4M addresses to be part in both address sets.
94\% of the addresses discovered via SRA probing, however, are unknown to the hitlist.
We will provide our data as new source to further improve the coverage of the hitlist service.

\vspace{-7pt}
\subsection{Comparison with IXP Flow Data}
We analyze one month of sampled (1:16k) IXP flow traffic.
During this period, we observe 2.5B~packets from 141M source and 87M destination addresses, resulting in a total of 198M~unique addresses, and 35M~addresses that appear as both source and destination. 
In terms of autonomous systems, we observe packets originating from 10k ASNs and targeting 11k ASNs, leading to 12k different ASes.
The flow traffic shows a bias towards a few, highly active ASNs that are responsible for more than 60\% of our packets (see \autoref{tbl:asn-allsources}).

Comparing our SRA dataset with the IXP flow data, which are aligned in time, we observe only 152k IPv6 addresses (0.2\% of our overall data) that are also visible in the flow data.

\vspace{-5pt}
\section{Routing Loops and Amplification}
\label{sec:loop-amples}

\begin{figure}
	\begin{subfigure}{0.48\textwidth}
    \centering
	  \includegraphics[width=\linewidth]{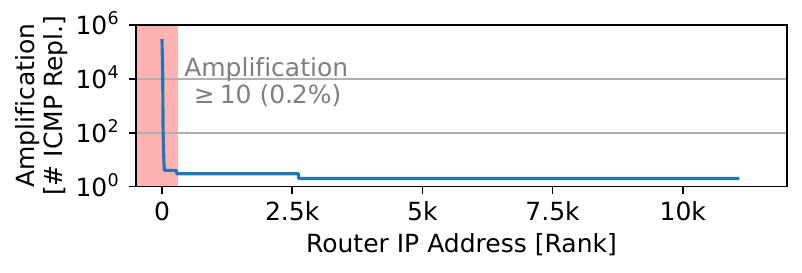}
	  \caption{Amplification factor per router IP address that sends more than one ICMP reply per request. For 98\% the amplification factor is $\leq 10$. Some few routers, however, amplify single requests by a factor of~>200k.}
	  \label{fig:results:max-ampl}
  \end{subfigure}
  \hfill
	\begin{subfigure}{0.50\textwidth}
    \centering
    \includegraphics[width=.96\linewidth]{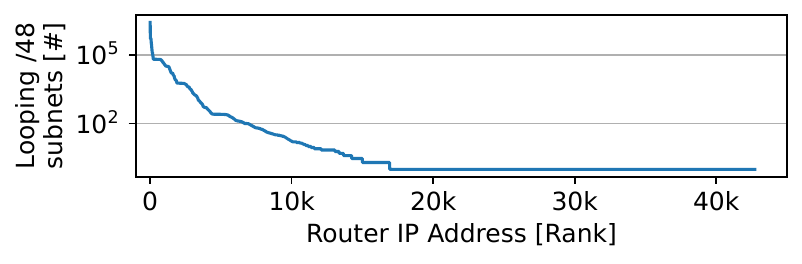}
    \caption{Frequency of /48 subnets that cause a routing loop per router. A few routers create routing loops to more than 100M /48 subnets.\\}
	  \label{fig:results:looping-subnets}
  \end{subfigure}
  \caption{Overview of routing loops and amplification factors.}
\vspace{-10pt}
\end{figure}

Routing loops~\cite{llzdl-finpd-21,mu-ilmsp-23,hmu-drwie-24} and amplification of looping packets~\cite{nkd-rlmad-22,gvr-obti-23} are a known problem in IPv6 deployments.
We confirm that both problems are still present.
They lack attention from the operator community.
During our active scans, we discovered a serious bug in the firmware of common router vendors that is triggered by routing loops and can lead to significant amplification.
A single ICMPv6 Echo request can result in more than 250,000~replies from the same router. 
We confirmed our observations with network operators and router vendors.
Routing loops, in conjunction with the amplification bug, can be exploited for denial-of-service attacks, posing a significant risk to network security.
Our SRA probing allows for careful, more complete discovery of this risk.

\paragraph{Reason for routing loops}
Routing loops are caused by incorrectly configured routes between a customer and provider such that packets to inactive subnets of provider aggregated address space assigned to the customer are sent back and forth between customer and provider until the IPv6~hop limit exceeds.
This misconfiguration can be fixed easily (details in \autoref{sec:loops-solution}).
What concerns us most is that some routers exponentially replicate the looping ICMPv6 Echos, which also leads to an exponentially increasing amplification of Time Exceeded messages.

\autoref{fig:results:max-ampl} visualizes the distribution of amplification factors gathered during our initial scan using a hop limit of 64~hops.
The magnitude of the amplification can be reduced by using even smaller hop limits. To reduce the impact of our scans on the network, we set the hop limit value to 64 for all follow-up measurements.

\paragraph{Prevalence of routing loops and amplification}
We use the results from our BGP /48 measurement to analyze the current deployment of routing loops in more detail.
We observe 141M~/48~subnets to trigger a routing loop.
These subnets affect 43k~router IP~addresses. 
The majority of the router IP~addresses (60\%) is responsible for a single subnet that triggers a routing loop, while some few routers connect more than 1M incorrectly configured /48 networks (see \autoref{fig:results:looping-subnets}).
Of the 41k~router IP addresses affected by routing loops, 11k are also affected by amplification (see \autoref{fig:results:max-ampl}).
These amplifying subnets highly concentrate in Brazil (23\%).
While 99.8\% of the amplification factors are below 10, a few routers amplify single ICMP Echo requests by a factor >100k (see \autoref{fig:results:max-ampl}).
These routers are primarily located in Germany and the US.
We provide more details on routing loops and amplification factors across countries in~\autoref{sec:dist-routingloops}.

\paragraph{Responsible disclosure and advice}
Packet floods harm performance of routers and links.
We argue that the measurement community should consider limiting IPv6~scans of inactive (or unknown) networks given the current state of deployment, as these scans could otherwise lead to loops and unintended consequences.
To monitor the current situation, we advise low frequency scans with small hop limits (\eg 64).
To improve the situation, we implemented a responsible disclosure policy and contacted 5340~network operators.
Originally, we observed routing loops in 141M~\texttt{/48}~subnets, which decreased in 263 ASes by a total of 7.7M loops until May, 2025.

\section{Discussion and Conclusion}
\label{sec:discussion}
\paragraph{Visibility}
We verified the visibility of all router IPv6 addresses that we found using SRA probing.
More than 70\% do not respond when queried directly. 
We argue that these addresses will rarely show up on hitlists because of their unresponsiveness.
Some of them could have been discovered by random probing and considering ICMPv6 error messages, which, however, will trigger more likely rate limiting.
Overall, reaching statistically reliable statements about the IPv6~population is still an open challenge, but we showed that reconsidering probing techniques may improve the stability of results significantly.

\paragraph{Stability and rate limiting}
Rate limiting is a key reason for instability of detected IPv6~addresses.
We showed that probing the SRA~address of a target subnet provides more stable results than random probing because SRA circumvents rate limiting of ICMPv6 error messages.
We observed, however, variations in the number of responses also for ICMPv6 Echo replies.
To what extent rate limiting techniques beyond those proposed in RFC~4443 are deployed should be part of future work.

\paragraph{Responsible scanning}
Active measurements of inactive address space may lead to significantly amplified traffic. %
Not all common measurement tools report about this unintended traffic, which then is only visible in raw packet captures.
We started to address these problems in collaboration with network operators and vendors.
We hope that our data will help to limit this unintended traffic in future research.

\vspace{-10pt}
\begin{acks}
We would like to thank our shepherd Philipp Richter and the anonymous reviewers.
This work was partly supported by the \grantsponsor{BMFTR}{Federal Ministry of Research, Technology and Space (BMFTR)}{https://www.bmbf.de/} within the projects IPv6Explorer (\grantnum{BMFTR}{16KIS1815}) and AI.Auto-Immune (\grantnum{BMFTR}{16KIS2333} and \grantnum{BMFTR}{16KIS2332K}).
\end{acks}

\label{lastbodypage}

\bibliographystyle{ACM-Reference-Format}
\bibliography{internet,security,rfcs,big-data,theory,own}

\appendix

\section{Ethics}

In this work, we found critical router configurations and a software bug of major router vendors, both are deployed on the IPv6 Internet.
We implemented a responsible disclosure policy by contacting all affected networks and router vendors.
First networks started to correct their configuration and router vendors started to fix their software.
Intentionally, we did not reveal networks and vendors in this paper.

We discovered these threats by active scanning, \ie sending standard ICMPv6 Echo requests with different hop limits.
This method is a common Internet measurement method and does not introduce ethical concerns.
As soon as we noticed the implications of our scans on some networks, we excluded the affected IP~prefixes from subsequent scans.
In addition, we excluded networks for which operators asked us to stop scanning---we received and processed a total of two opt-out requests.

\section{Artifacts}
\label{sec:artifacts}

All artifacts of this paper are publicly available.
These include
  \one all raw measurement data that we used to analyze SRA probing;
  \two all routing loop and amplification data after a clearance process;
  \two all data derived from post-processing, \ie data that substantiate our arguments and serve as input for our figures;
  \three measurement scripts;
  \four post-processing scripts.
All artifacts and details on how to use them are archived on \url{https://doi.org/10.5281/zenodo.17210254}.

Regular updates of our data are available on \url{https://ipv6-sra.realmv6.org}.

\section{Resolving Common Routing Loops}
\label{sec:loops-solution}
Internet routing loops often exist because a customer announces a covering prefix to its upstream provider but only maintains some routes to more specific prefixes of the overall announced address space.
In addition, this customer configures a default route via its upstream.
This makes parts of the address space that belong to the customer but are unused and locally unreachable accessible to the customer via the upstream.
To prevent routing loops (\eg triggered by downstream peers of the customer), the customer router needs to drop packets destined to the unused address space.
In many router implementations, this is achieved by using \emph{null routes}.
We now present necessary configurations to prevent routing loops on routers from Cisco and Juniper, two common router~vendors.

For example, a customer that has been assigned the prefix 2001:db8::/32 but only uses the parts 2001:db8::/34 and 2001:db8:8000::/34 can exclude all traffic to 2001:db8:4000::/34 and 2001:db8:c000::/34 from further forwarding as follows.

\paragraph{Cisco IOS}
\begin{verbatim}
  ipv6 route 2001:db8::/32 Null0
\end{verbatim}

\paragraph{Juniper Junos OS}
\begin{verbatim}
  set aggregate route 2001:db8::/32
\end{verbatim}

\noindent
Note that these configurations are examples.
The exact configuration syntax depends on the firmware~deployed.

\section{Overlap on AS Level}
\label{sec:overlap-asn}
\begin{figure}
  \centering
  \includegraphics[width=1\textwidth]{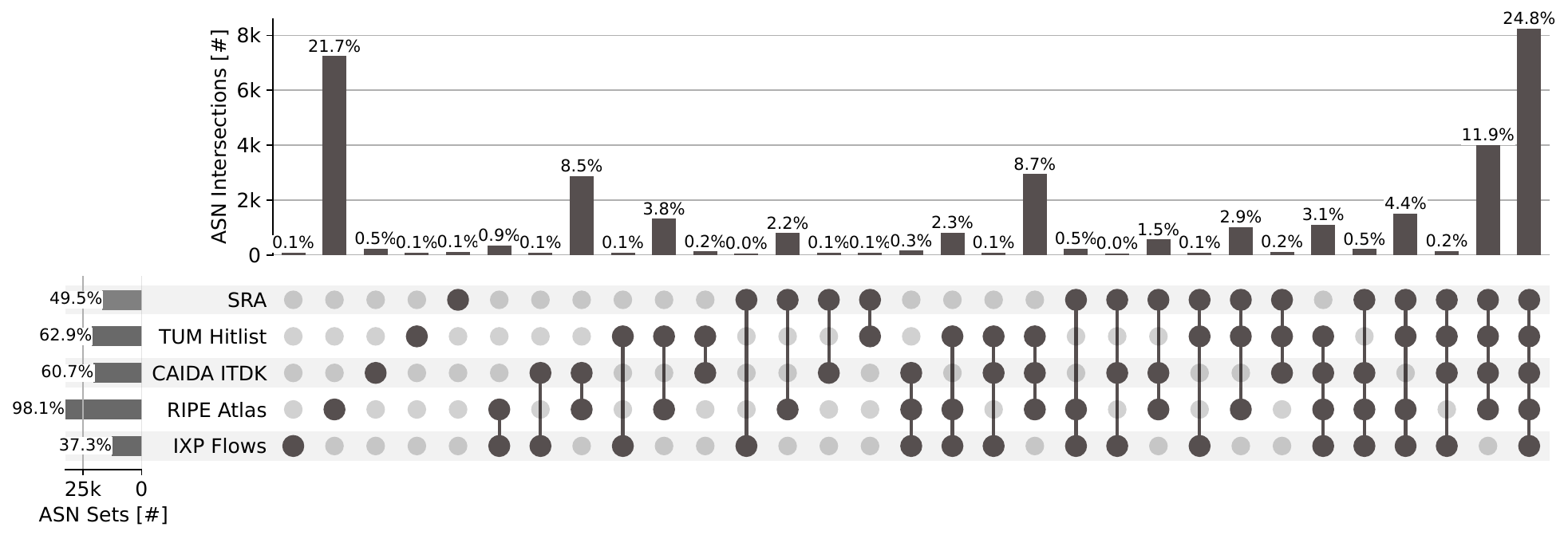}
  \caption{Overlap between all collected data sources on AS level.}
  \label{fig:overlap-asn-all}
\end{figure}
\autoref{fig:overlap-asn-all} shows the complete version of the UpSet plot presented in \autoref{fig:results:overlapASN}, \ie all combinations of intersections between the datasets we analyzed.
Additional combinations of intersections, however, do not change the overall picture of overlapping (and non-overlapping) ASNs that we discussed in \autoref{sec:sra-comparison}.

\section{Distribution of IP~Addresses Across Network Types}
\label{sec:network-types}
\begin{figure}
  \centering
  \begin{subfigure}{0.49\textwidth}
\includegraphics[width=1\textwidth]{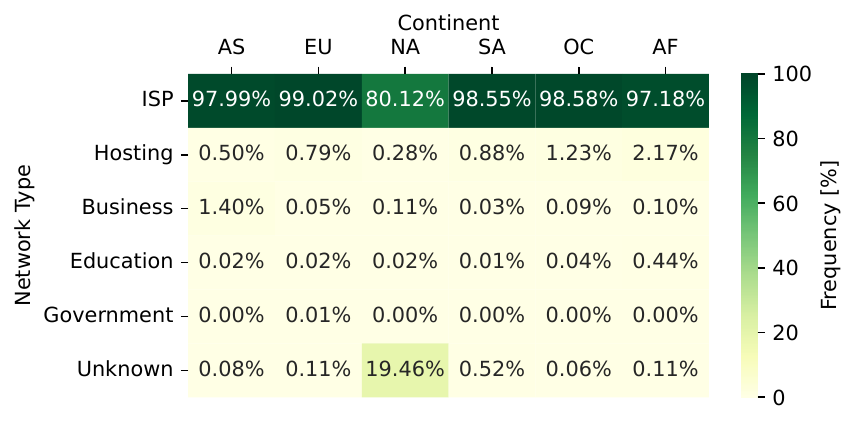}
  \caption{Router IP~addresses per continent discovered using SRA probing .}
  \label{fig:network-types-per-continent}
\end{subfigure}
\hfill
  \begin{subfigure}{0.49\textwidth}
    \includegraphics[width=1\textwidth]{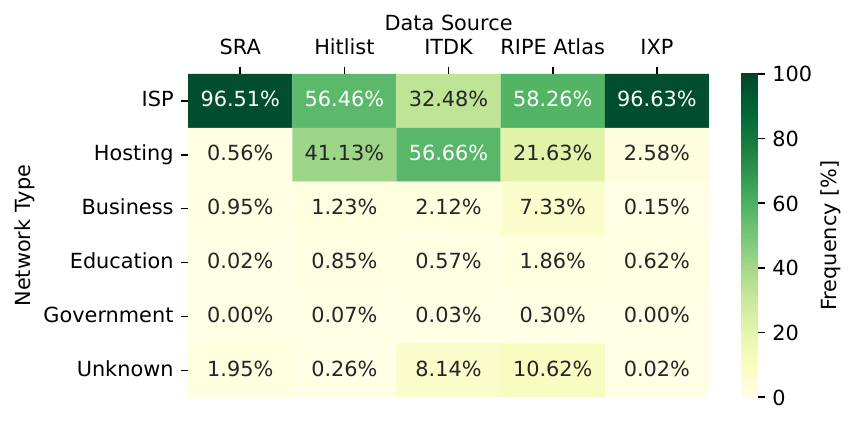}
    \caption{IP~addresses per network type, comparing our SRA and other datasets.}
    \label{fig:results:prefixstability}
  \end{subfigure}
  \caption{Distribution of IP~addresses per network type. Columns are sorted in decreasing order by absolute~values.}
  \label{fig:router-addresses-network-type}
\end{figure}

\autoref{fig:router-addresses-network-type} presents the distribution of discovered IP~addresses per network type, further grouped geographically for our SRA dataset and in comparison with other data sources.
Across all continents, the majority of router IP addresses we discovered using SRA probing belong to ISP networks (>80\%). 
Comparing SRA probing with other datasets, passive IXP flow data shows similar results, mainly including IP~addresses belonging to ISP networks (96\%).
All other sources contain, in addition to ISP networks, a significant fraction of IP~addresses linked to hosting networks.

\section{Distribution of Routing Loops and Amplification}
\label{sec:dist-routingloops}

\autoref{tbl:frq-routing-loops} shows the top 5 countries (Brazil, China, Czech Republic, Germany, Netherlands, and USA) that host the most /48 subnets affected by routing loops and amplification.
In total, we observe routing loops in more than 5k~autonomous systems, distributed across 155 countries.
54\% relate to infrastructure located in only five countries, with Brazil alone accounting for 26\%.
Interestingly, the routing loops in Brazil are distributed over 9k router IP addresses, whereas the routing loops in Germany, Czech Republic, and the Netherlands are distributed between 388 and 1.2k router~IP~addresses (see \autoref{tbl:rloops}).
In terms of maximum amplification factors, China is more relevant than the Netherlands (see \autoref{tbl:amplification}).

\begin{table}[h]
\setlength{\tabcolsep}{2pt}
  \caption{Top 5 countries hosting infrastructure that triggers routing loops and amplification.}
  \label{tbl:frq-routing-loops}
  \begin{subtable}[c]{0.4\textwidth}
    \caption{Top 5 countries ranked by the number of /48 subnets that trigger a routing loop.}
    \label{tbl:rloops}
    \begin{tabular}{lrrr}
\toprule
& \multicolumn{2}{c}{Looping /48 subnets} & Router\\
\cmidrule{2-3}
\cmidrule(lr){4-4}
      Country & [\#] & [\%] & IPs  [\#]\\
\midrule
      BRA &\num{ 37081970} & 26.22 & 9329 \\
      DEU &\num{ 13273666} & 9.39 & 1192 \\
      CZE &\num{ 10444197} & 7.39 & 881 \\
      USA &\num{ 7613551} & 5.38 & 4150 \\
      NLD &\num{ 7241713} & 5.12 & 388 \\
\bottomrule
\end{tabular}
  \end{subtable}
  \hfill
  \begin{subtable}[c]{0.52\textwidth}
    \caption{Top 5 countries ranked by the number of /48 subnets that trigger amplification.}
    \label{tbl:amplification}
\begin{tabular}{lrrrr}
\toprule
& \multicolumn{2}{c}{Ampl. /48 subnets}& \multicolumn{2}{c}{Router}\\
\cmidrule{2-3}
\cmidrule(lr){4-5}
  Country & [\#] & [\%] & IPs [\#] & Max. Ampl. [$\times$] \\
\midrule
  BRA & \num{4674687} & 23.35 & 2765 & 51 \\
  DEU & \num{477046} & 2.38 & 78 & \num{258399} \\
    USA & \num{247736} & 1.24 & 577 & \num{109385} \\
  CHN & \num{245546} & 1.23 & 4141 & 52 \\
  CZE & \num{212547} & 1.06 & 17 & 37 \\
\bottomrule
\end{tabular}
  \end{subtable}
\end{table}

\end{document}